\documentclass[published]{JHEP}

\usepackage{epsfig,amssymb}

\def\be{\begin{equation}}
\def\ee{\end{equation}}
\def\bea{\begin{eqnarray}}
\def\eea{\end{eqnarray}}

\preprint{FT-UAM-00-05, IFT-UAM/CSIC-00-05, hep-ph/0002076}

\title{Fermion production during preheating after hybrid inflation}

\author{Juan Garc{\'\i}a-Bellido\\
Departamento de F{\'\i}sica Te\'orica, C-XI, Universidad 
Aut\'onoma de Madrid, 28049 Madrid, Spain \\
Email: \email{bellido@delta.ft.uam.es}}
\author{Silvia Mollerach and Esteban Roulet\\
Departamento de F{\'\i}sica, Universidad Nacional de La Plata,
CC 67, 1900, La Plata, Argentina\\
Email: \email{mollerac@venus.fisica.unlp.edu.ar},
\email{roulet@venus.fisica.unlp.edu.ar}}

\received{February 7, 2000}

\abstract{
At the end of inflation, the coherent oscillations of the inflaton field
may resonantly amplify the long wavelength modes of both bosons and
fermions coupled to it. We study the resonant production of both kinds
of particles during preheating in a model of hybrid inflation. The
coherent time evolution of the inflaton and the Higgs fields after
inflation induce a very different production of fermions depending on
whether they are coupled to the Higgs or to the inflaton. For reasonable
values of the model parameters, the fermion production through
parametric resonance can be very efficient. We study the relative growth
of the fermion and boson energy densities during preheating in hybrid
models. During the initial stage of preheating, fermion production
dominates the relative energy density, while the exponential growth of
bosonic modes soon takes over.}


\begin{document}

\section{Introduction}

Cosmological inflation is an extremely efficient mechanism in diluting
any parti\-cle species or fluctuations. At the end of inflation, the
universe is empty and extremely cold, dominated by the homogeneous
coherent mode of the inflaton. Its potential energy density is converted
into particles, as the inflaton field oscillates  around the
minimum of its potential. These particles are initially very far from
equilibrium, but they interact among themselves (or decay) and thermal
equilibrium is achieved at a very large temperature. From there on the
universe expanded isoentropically, cooling down as it expanded, in the
way described by the standard hot Big Bang model. Recent developments in
the theory of reheating suggest that the decay of the inflaton energy
could have been explosive~\cite{KLS}, due to the coherent oscillations
of the inflaton zero mode, which induce its stimulated decay. The result
is a resonant production of particles in just a few inflaton
oscillations, in a process known as preheating~\cite{KLS}. The number of
particles produced in this way is exponentially large, which may account for
the extraordinarily large entropy, of order $10^{89}$, in our observable
patch of the universe today.

Preheating is not generic, it may occur in different models of
inflation~\cite{GKLS,GBL,GKS} but only under special circumstances.
Preheating strongly depends on the inflaton couplings to other fields,
as well as on the amplitude and frequency of the inflaton oscillations.
This explosive production occurs through parametric resonance of the
long wavelength modes of {\em any} field coupled to the inflaton, either
bosonic~\cite{KLS} or fermionic~\cite{GK}. Due to the parametric nature
of the resonance the particle production occurs for well defined
frequency bands. The difference between bosonic or fermionic preheating
is spectral: bosons can have large occupation numbers for a single mode,
while fermions saturate due of the Pauli exclusion principle, and
therefore a larger fraction of their energy can be transferred to higher
momentum modes. Also, the rate of growth of these two components 
is very different: while the boson energy density grows
exponentially~\cite{KLS}, that of fermions cannot~\cite{BHP}. As a
consequence, very soon bosons dominate the decay of the inflaton energy
if both kinds of particles are resonantly produced.  After a few
inflaton oscillations, the energy density in fermions and bosons may be
large enough to backreact on the inflaton oscillations and hence this
will eventually stop their resonant production~\cite{GKLS,GRTP}.

After preheating in chaotic inflation, the main part of the energy
density of the universe is in the coherent modes of bosons: the
inflaton, as well as any other parametrically amplified bosonic fields
with large occupation numbers, while a subdominant part may be in the
parametrically produced fermions. This state is very far from thermal
equilibrium (characterized by a decoherent ensemble with small
occupation numbers for all momentum modes) and, in fact, such a state
could be used to generate the required baryon asymmetry during
preheating, either via GUT baryogenesis~\cite{GUTB}, EW
baryogenesis~\cite{EWB} or leptogenesis~\cite{GRTP}. At the end, the
final thermalization of the universe occurs through the rescattering of
all the particles, which breaks the coherence of the bosonic fields. See,
for instance, Ref.~\cite{AS} for thermalization via fermionic modes.

In this paper we study the resonant production of both fermions and
bosons in a hybrid model of inflation~\cite{hybrid}. In these models it
is possible to choose, without fine-tuning, the masses and couplings of
the fields in such a way that the rate of expansion is negligible
compared to the masses involved~\cite{GBW}. In that case, preheating
occurs in a few inflaton oscillations after the end of inflation, before
the scale factor has grown by a single e-fold, and therefore we can
ignore the universe expansion. This reduces the problem to resonant
production of fermions in Minkowski space, for which there is a complete
analytic treatment~\cite{BHP}. We find that in the case of hybrid
inflation, contrary to what happens in chaotic inflation, it is possible
to produce, for certain couplings of the fermions to the inflaton field,
a larger contribution of fermionic modes than bosonic ones to the energy
density of the universe at the end of preheating.

We describe in Section 2 the hybrid model under consideration, with a
Higgs-type field coupled to the inflaton, and discuss the solution of
the classical equations of motion.  In Section 3 we study the pair
production of fermions in this model, for a non-zero coupling of
fermions to both the inflaton and the Higgs field. We analyze the growth
of the fermion energy density for different values of the couplings, as
well as for different fermion masses. In Section 4 we study the boson
production for both couplings to inflaton and Higgs, together with the
growth of the bosonic energy density.  In Section 5 we draw our
conclusions.

\section{Hybrid model}

We will consider a model of inflation with two fields, one that
slow-rolls down a potential, driving inflation, and another one with a
symmetry breaking potential that triggers the end of inflation, which we
will call the Higgs. Such a model is called hybrid inflation and was
proposed by Linde long ago~\cite{hybrid}. There are many particle
physics realizations of this general class of
models~\cite{CLLSW,GBLW,LR}. One particular potential for hybrid
inflation is given by the tree level potential~\footnote{See
Ref.~\cite{Lyth} for a discussion of loop-corrections to this potential
during inflation.}
\begin{equation}\label{potential}
V(\phi,\sigma)=\frac{M^4}{4\lambda}-\frac{1}{2}M^2\phi^2+
\frac{1}{4}\lambda\phi^4 + \frac{1}{2}g^2\phi^2\sigma^2 +
\frac{1}{2}m^2\sigma^2\,.
\end{equation}
During hybrid inflation, the inflaton field $\sigma$ evolves along the
Higgs valley at $\phi=0$. As soon as the Higgs acquires a negative mass
term, it triggers the end of inflation.  That is, in less than one
e-fold the two fields start oscillating around their common absolute
minimum $(\phi=v, \sigma=0)$, where $v=M/\sqrt{\lambda}$ is the Higgs
vacuum expectation value (vev). The end of inflation occurs when the
mass of the Higgs vanishes, i.e. at $\sigma=\sigma_c\equiv M/g$.

By defining $y\equiv \sigma/\sigma_c$ and $x\equiv\phi/v$, we can write
the potential (\ref{potential}) as
$$V(x,y) = V_0[(1-x^2)^2 + 2x^2 y^2 + 2\gamma y^2]\,,$$ where $V_0 =
M^4/4\lambda$, and $\gamma = \lambda m^2/g^2M^2$ \ is a constant that is
constrained to be small by the amplitude of the microwave background
(CMB) anisotropies. It does not play any significant role after
inflation and we will neglect it here. Also, the rate of expansion can
be made relatively small in hybrid inflation (as long as $M\ll gM_{\rm P}$)
and still satisfy the CMB
constraints, see Ref.~\cite{GBW}. Therefore, we will ignore the rate of
expansion of the universe here. We will also concentrate in the regime 
 $\lambda\gg g^2$, for which the Higgs evolves along the minimum of the
potential, following the inflaton oscillations.

We can then write the evolution equations after inflation, 
 redefining the time unit as $\tau\equiv \bar M t$ (with $\bar M =
g M/\sqrt{\lambda}$), as:
\begin{eqnarray}\label{evolution}
y''+(1-y^2)y &=& 0 \,,\\
1-y^2 &=& x^2 \,.
\end{eqnarray}
For initial conditions $y(0)=y_0$ and $y'(0)=0$, we find the solution
\begin{equation}\label{solution}\begin{array}{rl}
\hspace{2.5cm}y=y_0\,{\rm cd}\,(u|m)\,, &\\[2mm]
u=\tau\sqrt{a_2/2}\,,\hspace{1.6cm}& m={a_1/a_2}\,,\\[2mm]
a_1=y_0^2\,,\hspace{2.5cm}& a_2=2-y_0^2\,,
\end{array}\end{equation}
where ${\rm cd}\,(u|m)$ is the Jacobi elliptic function.
This solution is periodic in the $\tau$ variable, with period
\begin{equation}\label{period}
T=4\sqrt{2\over a_2}\,K(m)\,,
\end{equation}
with $K$ the elliptic integral. We have plotted the evolution of both
the normalized Higgs and the inflaton after inflation in
Fig.~\ref{fig1}. Throughout the paper we will take $y_0 = 0.9999$. The
corresponding period of the inflaton oscillations is $T \simeq 29.97$.
Although this period increases with $y_0$ approaching one, the mean
energy density transferred to fermions coupled to the inflaton is nearly
unchanged (provided that $y_0$ is close to one). If the fermions are
coupled to the Higgs, the results are more sensitive to the value of
$y_0$ for strong coupling, as we will discuss later.

\FIGURE{
\epsfig{file=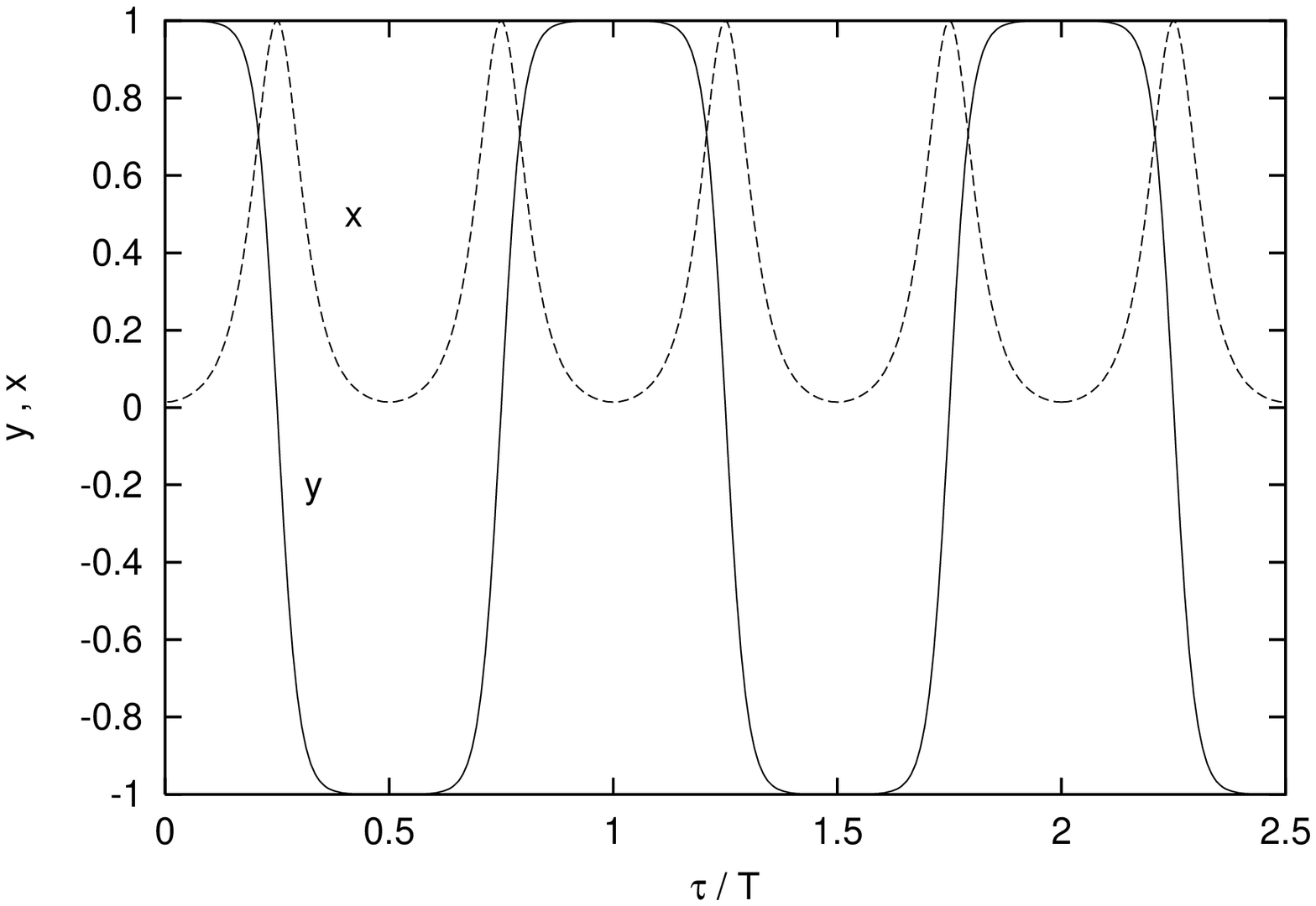,width=8truecm,angle=-90}
\caption{The time evolution of the inflaton and 
Higgs fields after inflation, for initial condition $y_0=0.9999$.}
\label{fig1}
}

\section{Pair production of fermions}

The oscillations of the inflaton and Higgs fields at the end of
inflation trigger the explosive production of particles that couple to
either or both of these fields. The production of bosons in hybrid 
inflation has been studied in Ref.~\cite{GBL,BKS}. Here we will analyze 
the fermion production in this model and in the next Section will  
elaborate further on the bosonic case in order to compare the two results.

We will consider first the coupling of fermions
to the inflaton $\sigma$, with coupling $h_1\sigma\bar\psi\psi$, and a
possible mass term $m_\psi\bar\psi\psi$.  Then we will consider the
coupling of fermions to the Higgs, with coupling
$h_2\phi\bar\psi\psi$. When the symmetry gets broken this last coupling
will give the fermions a mass 
through the non vanishing vev of the Higgs.

\subsection{Coupling to the inflaton}

Let us consider here a fermionic field $\psi$ satisfying the Dirac
equation
\begin{equation}
\left(i \gamma^\mu \nabla_\mu - h_1 \sigma (t) - m_\psi\right)\psi=0.
\label{dirac}
\end{equation}
The solutions are more easily obtained using an auxiliary field 
$X(\vec{x},t)$, such that $\psi = \left(i \gamma^\mu \nabla_\mu + 
h_1 \sigma (t) + m_\psi\right) X$. Decomposing it as $\exp(i
\vec{K} \cdot \vec{x}) X_K(t) R_r$, with $R_r$ eigenvectors of
$\gamma^0$ with eigenvalue $+1$, we can write the
equation of motion for fermion modes $X_k$ as~\cite{GK}
\begin{equation}\label{mathieu}
X_k''+\Omega_k^2X_k-i\sqrt{q_1}\,y'X_k=0\,,
\end{equation}
where 
\begin{eqnarray}\label{omega}
\Omega_k^2(\tau)&=&k^2+(\sqrt{q_1}\,y(\tau)+\bar m_\psi)^2\,,\\
q_1&\equiv&{\lambda h_1^2\over g^4}\,,\\
\bar m_\psi&\equiv&{m_\psi/\bar M}\,
\end{eqnarray}
and we have rescaled $k\equiv K/\bar M$. 
We will here display the results for $q_1$ between 1 and $10^6$. Notice
that if we were to take e.g. $\lambda=1$ and $g=0.01$  this will lead to
$q_1=10^8h_1^2$, and hence the values of $q_1$ considered  would
correspond to $h_1$ between $10^{-4}$ and 0.1.

We take the  initial conditions corresponding to positive frequency
plane waves at $\tau<0$:
\begin{eqnarray}\label{initcond1}
X_k(0)&=&[2\Omega_k(\Omega_k+\sqrt{q_1}\,y_0+\bar m_\psi)]^{-1/2}\,,\\
X_k'(0)&=&-i\Omega_kX_k(0)\,.\label{initcond2}
\end{eqnarray}

The quantity of interest to us is the fraction of the total energy,
$\rho_{\rm total}=V_0$, which is transferred into fermions, 
\begin{equation}\label{ratiorhoinf}
{\rho_{_F}(\tau)\over \rho_{\rm total}}= 
h_1^2\,{2 N\over\pi^2 q_1}\int dk\,k^2\,
\Omega_k(\tau)  n_k(\tau)\,,
\end{equation}
with the number of fermion-pair degrees of freedom $N=1$ or 2 for
Majorana or Dirac fields, respectively.
The occupation number (for fermion pairs), $n_k$,  can be calculated 
as~\cite{GMM}
\begin{equation}\label{occno}
n_k(\tau)=\frac{1}{2}-{k^2\over \Omega_k}\
{\rm Im}\,[X_k{X'_k}^*]-{\sqrt{q_1}\,y+\bar m_\psi \over 2\Omega_k}\,.
\end{equation}
Note that $n_k(0)=0$, thanks to the initial conditions~(\ref{initcond1})
and (\ref{initcond2}). Also, it is easy to see that the occupation
number of fermion pairs is always smaller than one,\footnote{This is
related to the fact that $n_k = |\beta_k|^2 = 1-|\alpha_k|^2\leq1$,
for fermions, in terms of the Bogoliubov transformation coefficients
($\alpha_k, \beta_k$), see Ref.~\cite{GMM}.} as expected.

\FIGURE{
\epsfig{file=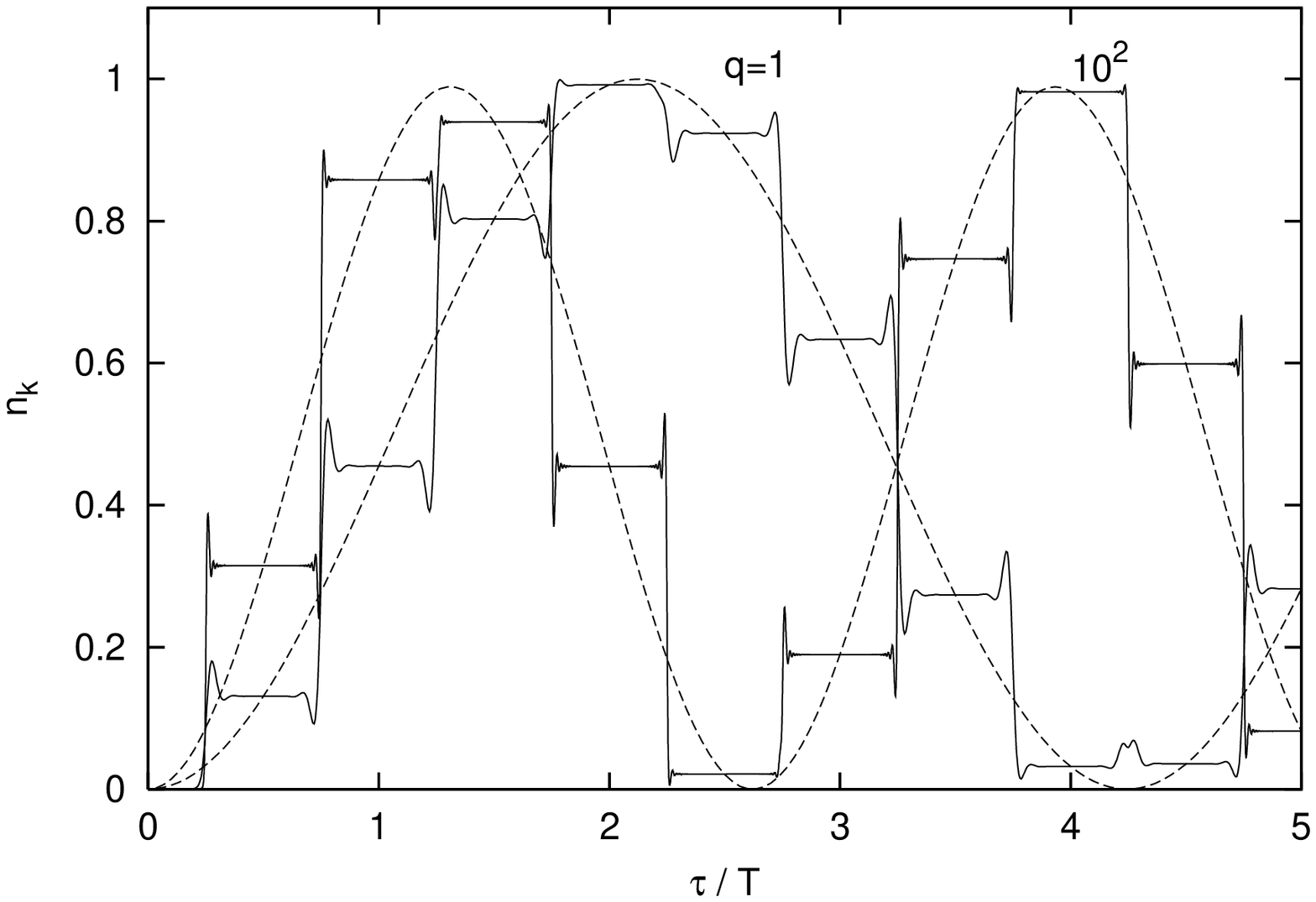,width=8truecm,angle=-90}
\caption{ The time evolution of the fermion occupation number $n_k$
(continuous line) and the averaged particle number $\bar n_k$ (dashed
line), for values of $k^2=0.51 $ and 1.4, corresponding to the first 
peaks of $F_k$
for $q_1=1$ and 100 respectively.}
\label{nki}
}

Particle production through parametric resonance occurs at those
moments when the adiabaticity condition d$\Omega_k/$d$\tau<\Omega_k^2$ is
violated, see Ref.~\cite{KLS}. 
For fermion production this will occur  in general 
whenever the effective fermion
mass $\bar m_\psi+\sqrt{q_1}y$ approaches zero. 
This is illustrated in Fig.~\ref{nki}, where the continuous
lines show the evolution of the exact occupation number for massless
fermions and two values of the $q_1$ parameter. 
The occupation
numbers $n_k$ (solid lines) have 
jumps every quarter and three quarters of the
inflaton period, corresponding to the times when the inflaton value
vanishes, $y=0$ (see Fig.~\ref{fig1}), and therefore when the effective
fermion mass also vanishes. 

To evaluate Eq.~(\ref{ratiorhoinf}) using $n_k(\tau)$ from
Eq.~(\ref{occno}) and the numerical solutions of (\ref{mathieu}) is
however quite demanding, so that it is convenient to use the approximate
analytical method developed in \cite{GMM,GK}. This method exploits the
periodicity of $y$ and allows to obtain a smooth function
\begin{equation}\label{barnk}
\bar n_k(\tau)=F_k\sin^2(\nu_k\tau)\,,
\end{equation}
which coincides with $n_k(\tau)$ for $\tau=nT$.  Hence, $\bar n_k(\tau)$
gives an approximate expression for $n_k(\tau)$ without its fine (spiky)
details.  The approximate solution (dashed lines in Fig.~\ref{nki})
follows the overall oscillations of the occupation number, matching the
exact results at every inflaton period (and also at every half period in
this case).

\FIGURE{\epsfig{file=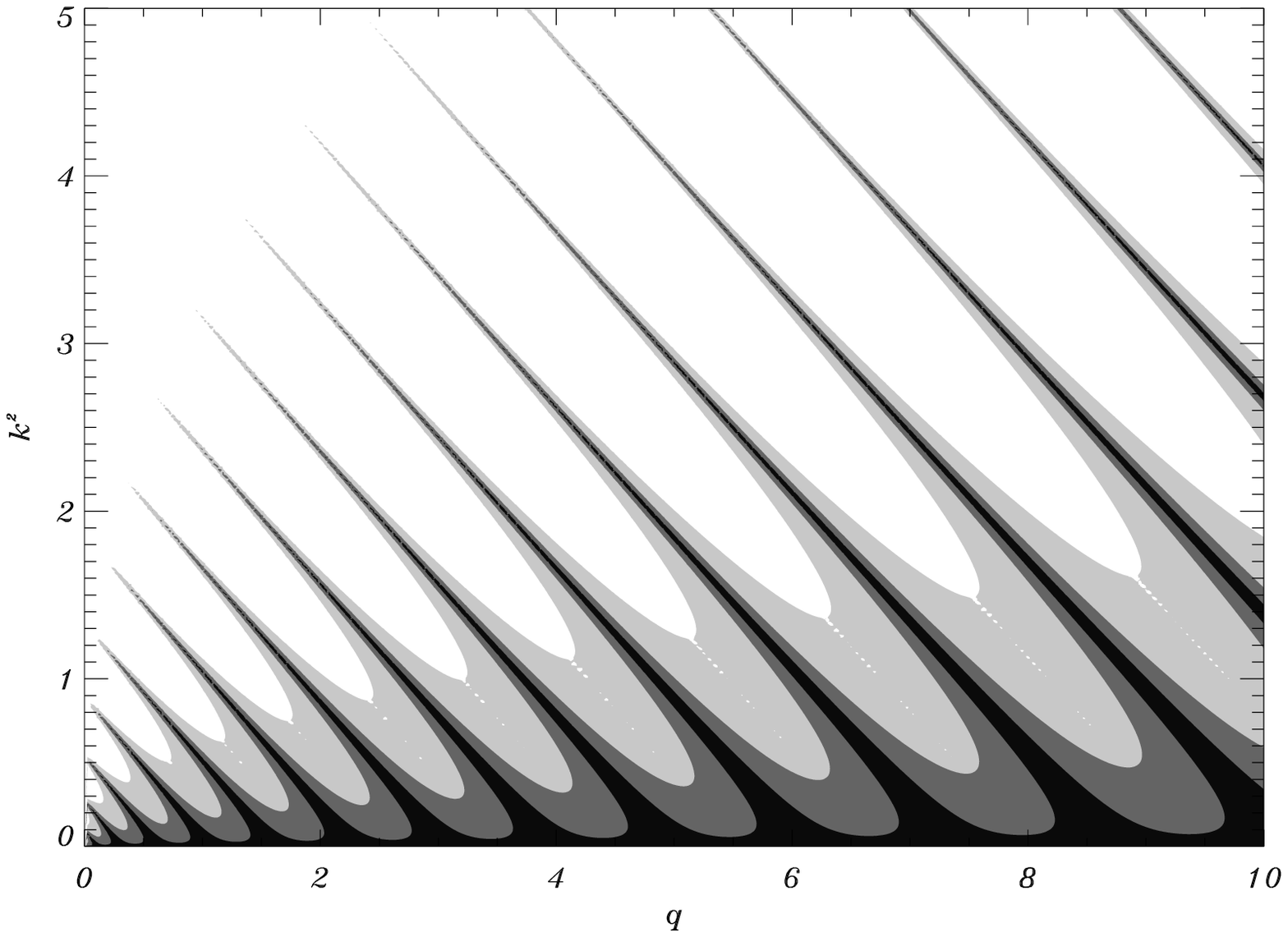,width=12truecm,angle=0}
\caption{ The instability chart for fermions coupled to the
inflaton. Contours correspond to equipotential values of 
$F_k$=0.1, 0.5 and 0.9  in the plane $(q_1, k^2)$ from lighter to darker.}
\label{insti}
}

The advantage of using $\bar n_k(\tau)$ is that it has a simple temporal
behaviour, while its $k$ dependence can be obtained from the knowledge
of the functions $F_k$ and $\nu_k$, where
\begin{eqnarray}\label{Fk}
F_k&=&{k^2({\rm Im}\,X_k^{(1)}(T))^2\over \Omega_k^2(T)
\sin^2(\nu_kT)}\,,\\[2mm]
\cos(\nu_k T)&=&\pm\,{\rm Re}\,X_k^{(1)}(T)\,.\label{nuk}
\end{eqnarray}
and $X_k^{(1)}$ satisfies the same equation (\ref{mathieu}) with the
initial condition $X_k^{(1)}(0)=1$, ${X_k^{(1)}}'(0)=0$. Therefore, to
obtain $\bar n_k (\tau)$ we only need to solve Eq.~(\ref{mathieu})
during one inflaton period.  The two signs in Eq.~(\ref{nuk}) correspond
to two possible functions ${\bar n}_k$, which oscillate with the same
amplitude but different frequency, and both match $n_k$ at every period
of the inflaton oscillation. The best approximation to $n_k(\tau)$ is
given by the one with smaller frequency $\nu_k$ for the $k^2$ values
where $F_k$ is maximum (i.e. for the momenta contributing significantly
to $\rho_{_F}$), which in this case corresponds to taking the minus sign
in Eq.~(\ref{nuk}). This is the function plotted in Fig.~\ref{fknuk}.
Notice that the maximum value of $\nu_k$ is $\pi/T\simeq 0.105$, as can
be deduced from Eq.~(\ref{nuk}).

\FIGURE{\epsfig{file=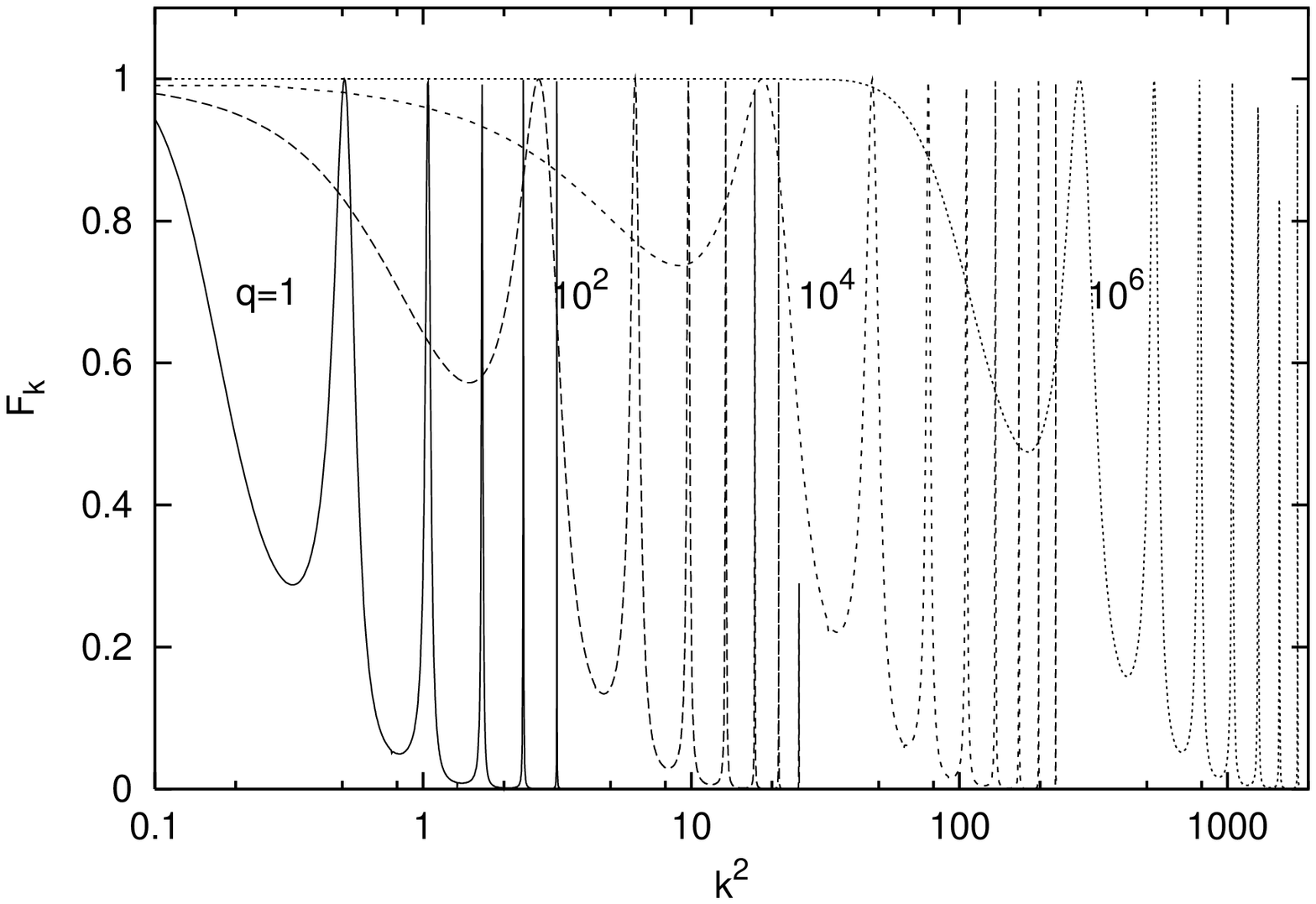,width=8truecm,angle=-90}
\epsfig{file=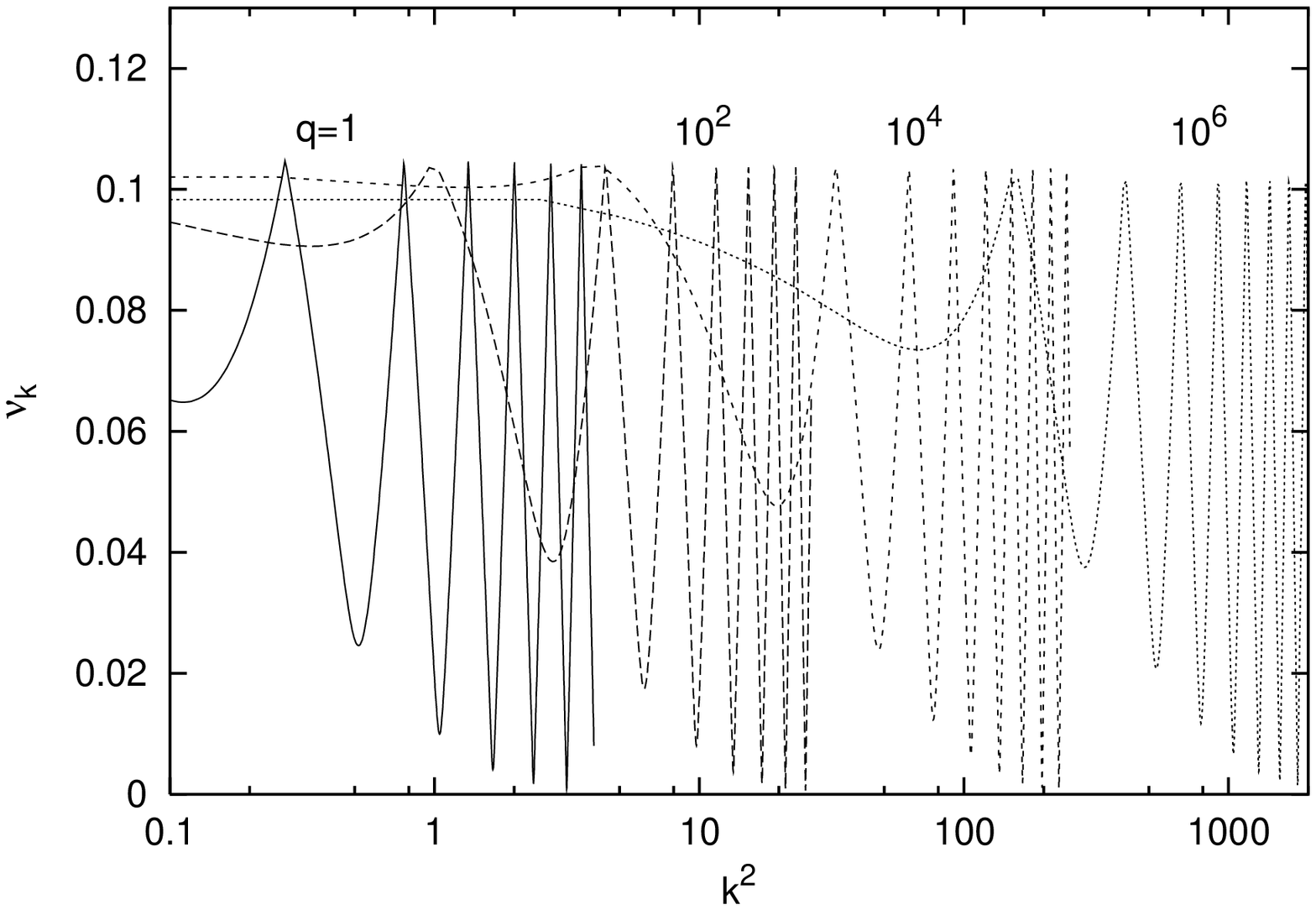,width=8truecm,angle=-90}
\caption{ The fermion (envelope) spectrum $F_k$ (upper panel)
and $\nu_k$ (lower panel) as a function of 
$k^2$, for fermions coupled to the inflaton, with 
$q_1=1, 10^2, 10^4, 10^6$.}
\label{fknuk}
}

We show in Fig.~\ref{insti} the instability chart for massless fermions
coupled to the inflaton. It displays the contours in the $(q_1, k^2)$
plane of equal $F_k$ values. Fermions are mainly produced with momenta
in the darker regions, corresponding to maxima of $F_k$. We see that the
bands get narrower with increasing $k^2$ for a given $q_1$ value, and
after several bands they shrink to a negligible width.

The upper panel in Fig.~\ref{fknuk} shows the behaviour of $F_k$ for
different $q_1$ values extending up to $q_1 = 10^6$. These correspond to
cuts in the instability chart (Fig.~\ref{insti}) at a fixed value of
$q_1$. The maximum momentum for which the bands are sizeable grows as
$k_{\rm max} \sim q_1^{1/4}$ for fermions coupled to the
inflaton.\footnote{Note that the same behaviour was found in
Ref.~\cite{GK} for chaotic inflation. However, a different behaviour,
$k_{\rm max} \sim q^{1/3}$, was found in Ref.~\cite{GRTP} and was
attributed, in their case, to the redshift of the modes during the
particle production process.} The lower panel of Fig.~\ref{fknuk} shows
$\nu_k$, as a function of $k^2$, for the same values of $q_1$. Note that
the maxima of $F_k$ correspond to local minima of~$\nu_k$.

\FIGURE{\epsfig{file=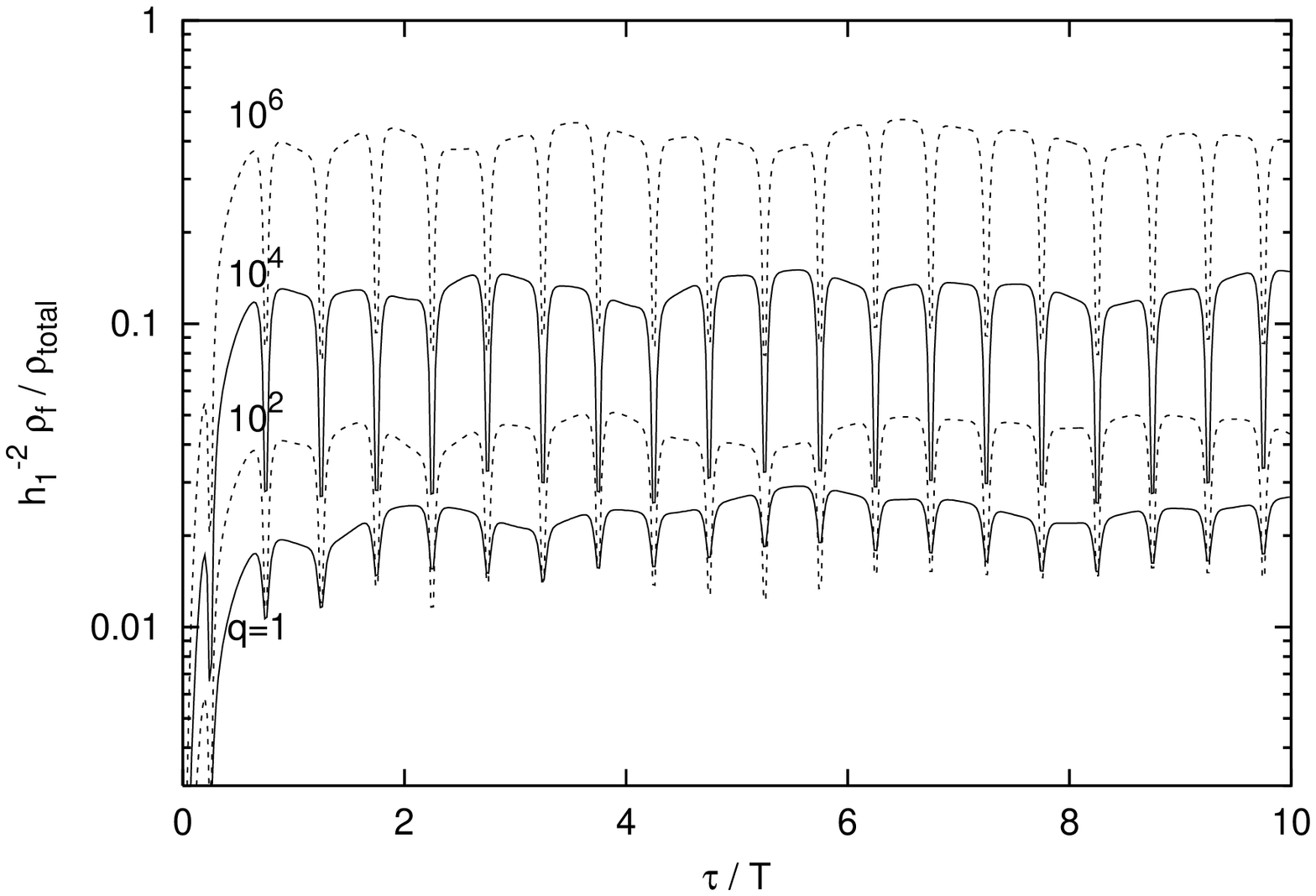,width=8truecm,angle=-90}
\caption{ The fraction of the total energy transferred to fermions
coupled to the inflaton, as a function of time, for $q_1 = 1, 10^2,
10^4, 10^6$.}
\label{rhoivst}
}

The fraction of the total energy transferred to the fermions then 
finally results
\begin{equation}\label{rhoF}
{\rho_{_F}(\tau)\over \rho_{\rm total}}= 
h_1^2\,{2 N\over\pi^2 q_1}\int dk\,k^2\,\Omega_k(\tau)
F_k \sin^2(\nu_k\tau)\,.
\end{equation}

Its time evolution is shown in Fig.~\ref{rhoivst}, for different values of
the coupling parameter $q_1$. We see that after the first oscillation of
the inflaton this fraction already reaches its asymptotic value and then
fluctuates around it. The asymptotic value $\rho_{_F}/\rho_{\rm total}$
scales as $q_1^{1/4} h_1^2$. 
Moreover, for  large values of $q_1$ a significant fraction of the
inflaton energy can be transferred into fermions (as long as $h_1$ is not 
too small). We have checked that the final density is
insensitive to the initial value of the inflaton field $y_0$ as long as
$|y_0-1| \leq 10^{-2}$. This can be simply understood since, for $m_\psi=0$,
the production of fermions takes place as the inflaton crosses through
$y=0$.

\FIGURE{\epsfig{file=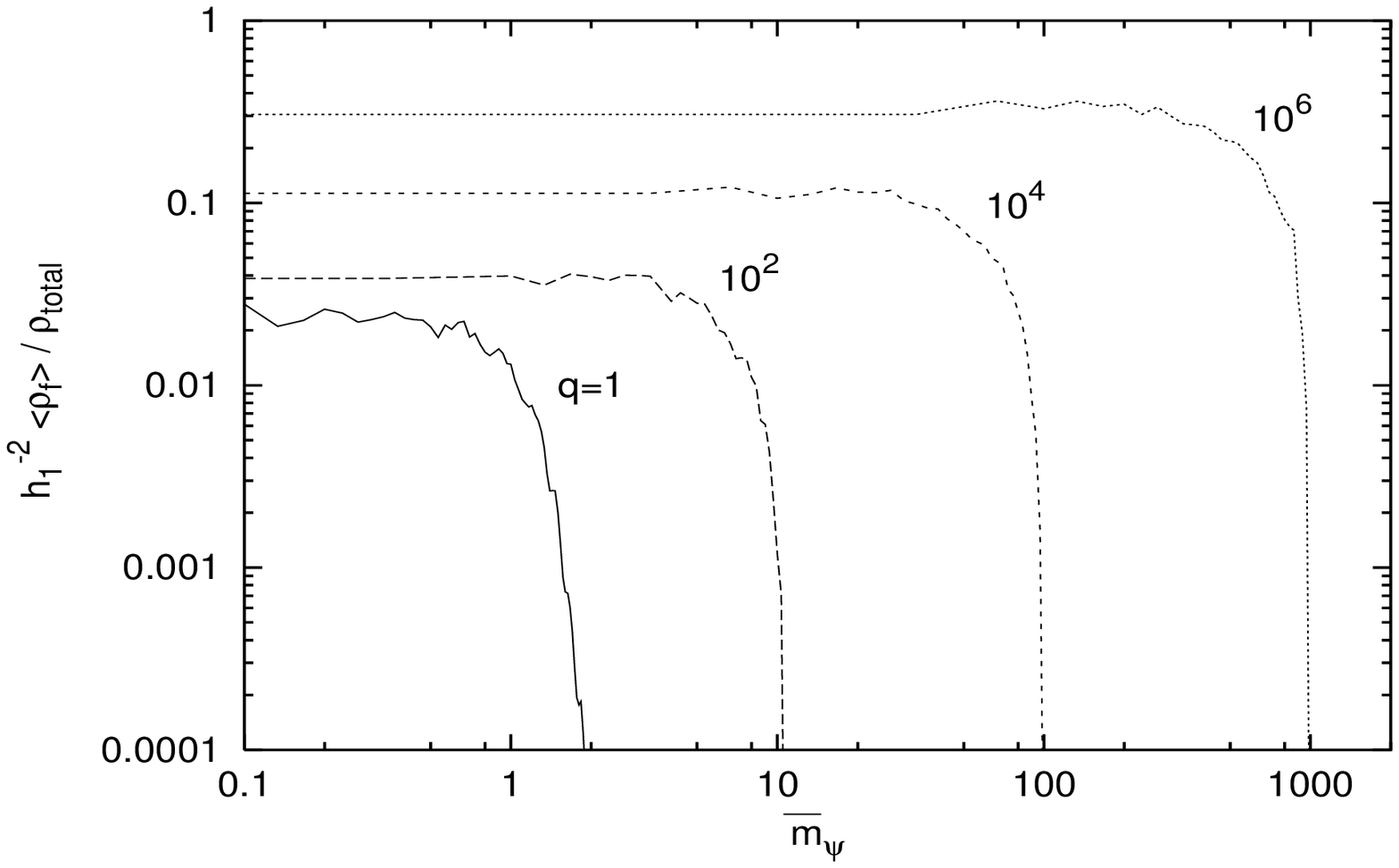,width=9truecm,angle=-90}
\caption{ The fraction of the average energy density in fermions, 
$h_1^{-2}\langle\rho_{_F}\rangle/\rho_{\rm total}$ as a
function of the fermion mass $\bar m_\psi$ (averaged between the eighth
and tenth inflaton periods).}
\label{rhomass}
}

If fermions are massive, the inflaton energy is transferred to the
fermions with nearly the same efficiency as for massless fermions up to
a maximum cut off value of the fermion's mass, above which it
drastically drops. Fig.~\ref{rhomass} shows this behaviour for different
values of the $q_1$ parameter. The cut off value of the mass goes like
$\ \bar m_{\psi,\rm cutoff} \sim q_1^{1/2}$.~\footnote{The same
behaviour was found for chaotic inflation in Ref.~\cite{GRTP}.} The
reason is simple, particle production occurs whenever the effective
fermion mass vanishes, $m_{\rm eff} = \bar m_\psi + \sqrt{q_1}\,y(\tau)
= 0$. For small values of $q_1$, even a small (positive) $\bar m_\psi$
will prevent $m_{\rm eff}$ from vanishing as the inflaton $y(\tau)$
oscillates. However, as we increase $q_1$, larger bare masses are still
allowed for particle production. The largest (negative) value of
$y(\tau)\simeq-1$ gives the cutoff mass.  Recalling that the actual mass
of the fermion is $m_\psi=(g/\sqrt{\lambda})\bar m_\psi M$, and that we
are working on the regime $M<g M_{\rm P}$, we see that this still allows
for quite large fermion masses to be produced.

\subsection{Coupling to the Higgs}

\FIGURE{\epsfig{file=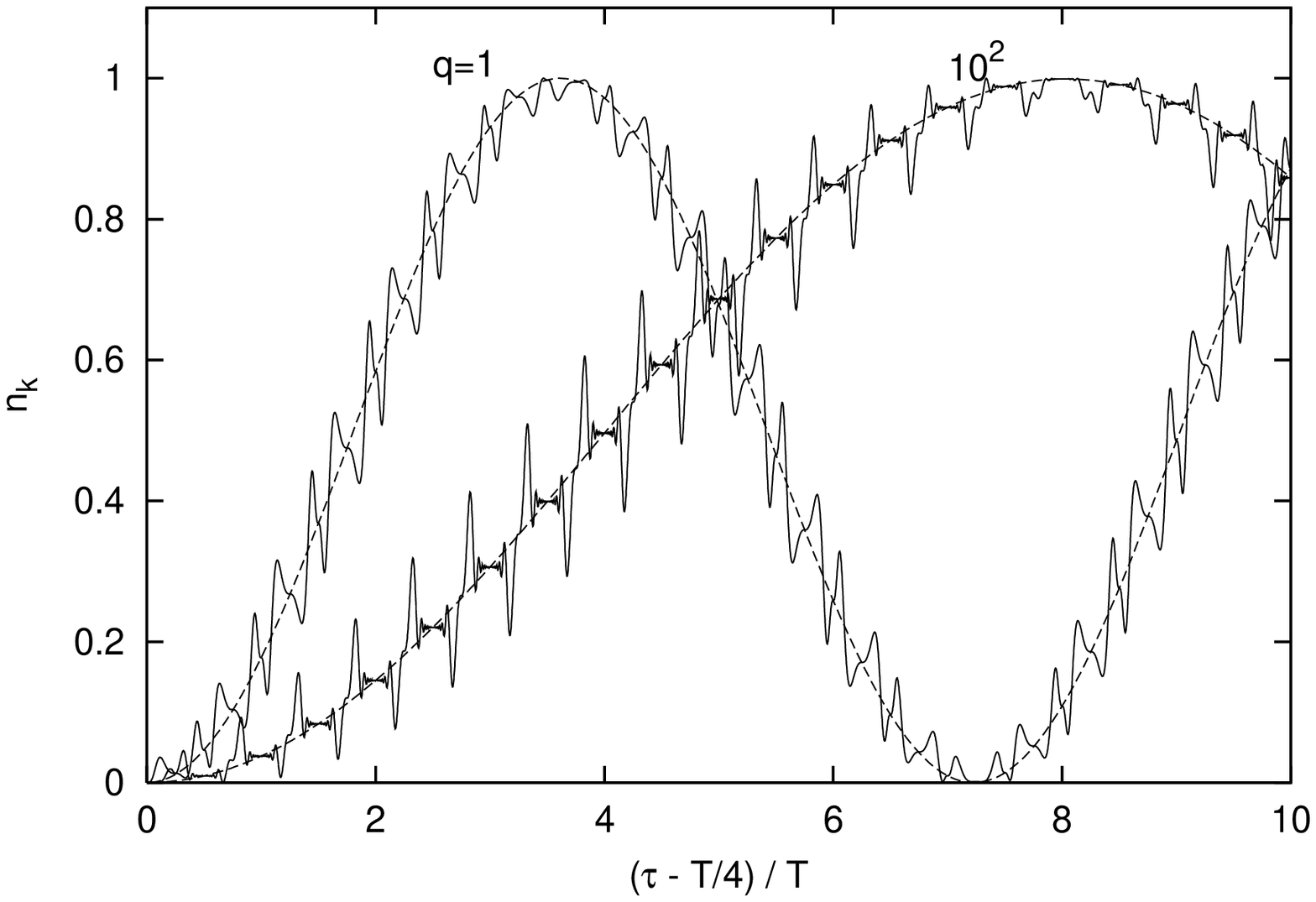,width=8truecm,angle=-90}
\caption{ The time evolution of the fermion occupation number $n_k$
(continuous line) and the approximating function $\bar n_k$ (dashed
line), for values of $k^2=0.235 $ and  0.371, corresponding to 
peaks of $F_k$
for $q_2=1$ and 100 respectively.}
\label{nukh}
}

We will consider now the coupling of fermions to the Higgs, and
study their pair production. The associated Mathieu equation is
analogous to (\ref{mathieu}), but with $x(\tau)=+\sqrt{1-y(\tau)^2}$ 
in the place of $y(\tau)$, i.e. 
\begin{equation}\label{matHiggs}
X_k''+\Omega_k^2 X_k-i\sqrt{q_2}\,x' X_k=0\,,
\end{equation}
where $\Omega_k^2=k^2+q_2\,x^2$, \ $x'=-y'\,y/x$, and $q_2=h_2^2/g^2$.

\FIGURE{\epsfig{file=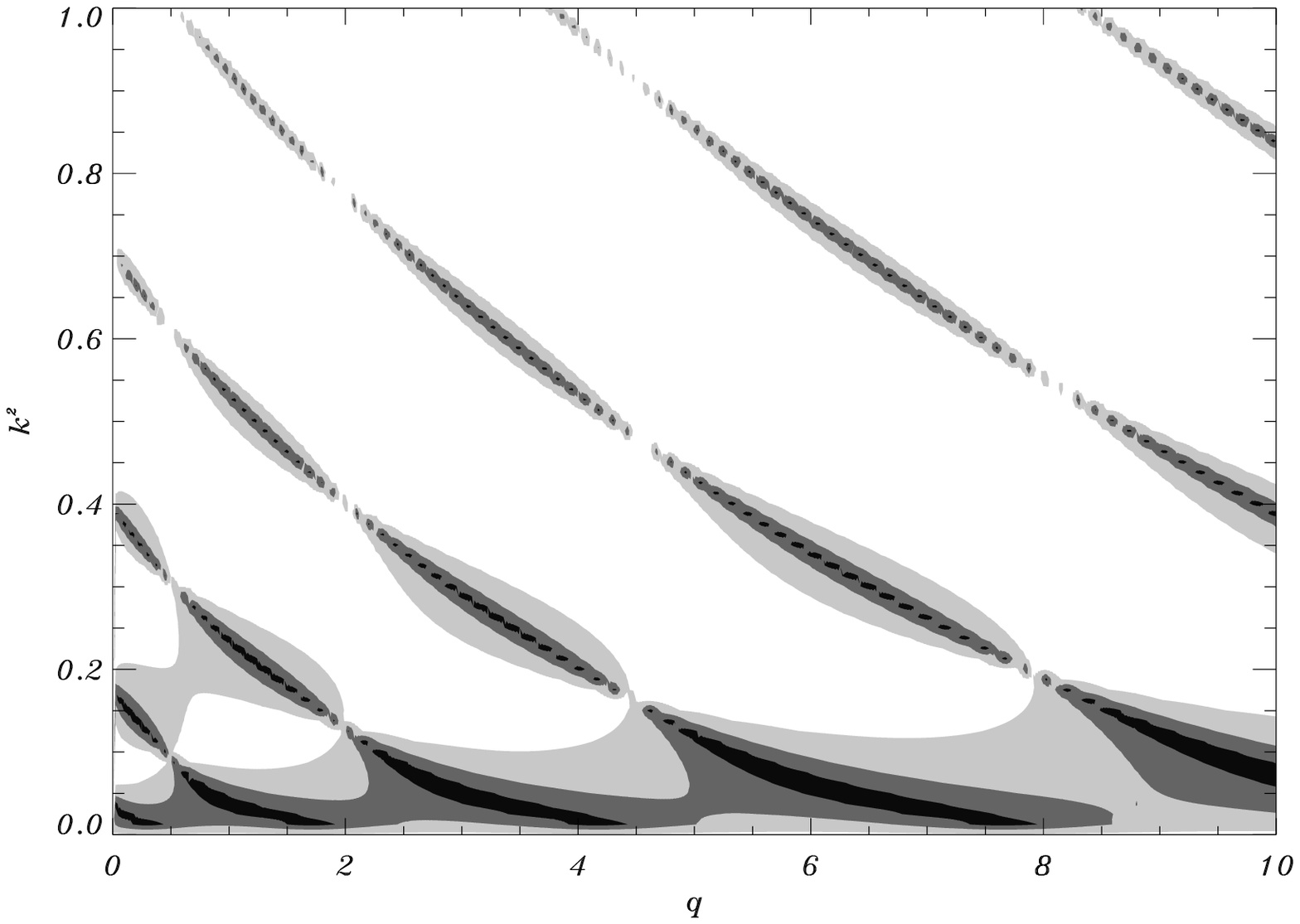,width=12truecm,angle=0}
\caption{ The instability chart for fermions coupled to the
Higgs. Contours correspond to equipotential values of 
$F_k$=0.1, 0.5 and 0.9  in the plane $(q_2, k^2)$.}
\label{insth}
}

The initial conditions are given by (\ref{initcond1}) with $y_0\to~x_0$
and we take the bare mass $\bar m_\psi=0$, since we are assuming that
the fermion acquires a mass through the Higgs mechanism.

The occupation number is given by
\begin{equation}\label{partno}
n_k(\tau)=\frac{1}{2} - {k^2\over \Omega_k}\
{\rm Im}\,[X_k{X'_k}^*] - {\sqrt{q_2}\,x \over 2\Omega_k}\,.
\end{equation}
Particle production again occurs periodically. In this case, the jumps
in the occupation number take place for every period and half period of
the inflaton, i.e.  at $\tau= n T/2$.  These correspond to the times
when the Higgs approaches zero, and thus when the effective mass of the
fermion is at a minimum. An approximate expression $\bar n_k(\tau)$ can
be constructed as in the previous case of fermions coupled to the
inflaton. However, a function that matches $n_k(\tau)$ at the times when
it has the spikes would not generally be a good approximation to the
$n_k(\tau)$. Thus we have performed a shift of a quarter of period in
the initial time, so that $\bar n_k (\tau)$ matches the value of
$n_k(\tau)$ at $\tau = (2n+1)T/4$, i.e. where $n_k(\tau)$ has the plateau,
and this gives a very good approximation as it is shown in
Fig.~\ref{nukh}. Starting with $n_k(\tau)=0$ at $\tau=T/4$ does not
affect the results after a few inflaton oscillations, since the density
rapidly saturates to its asymptotic value.

\FIGURE{\epsfig{file=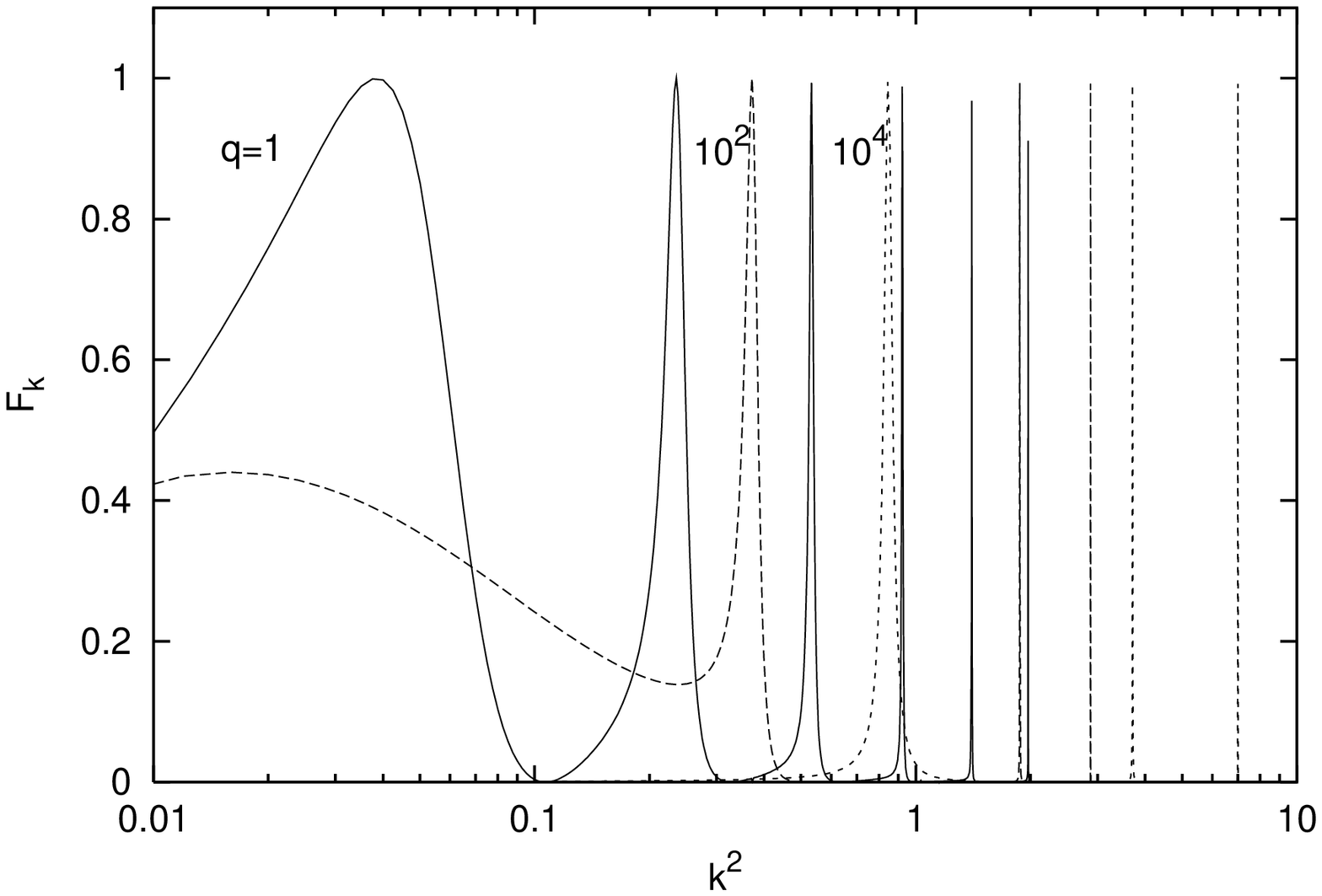,width=8truecm,angle=-90}

\epsfig{file=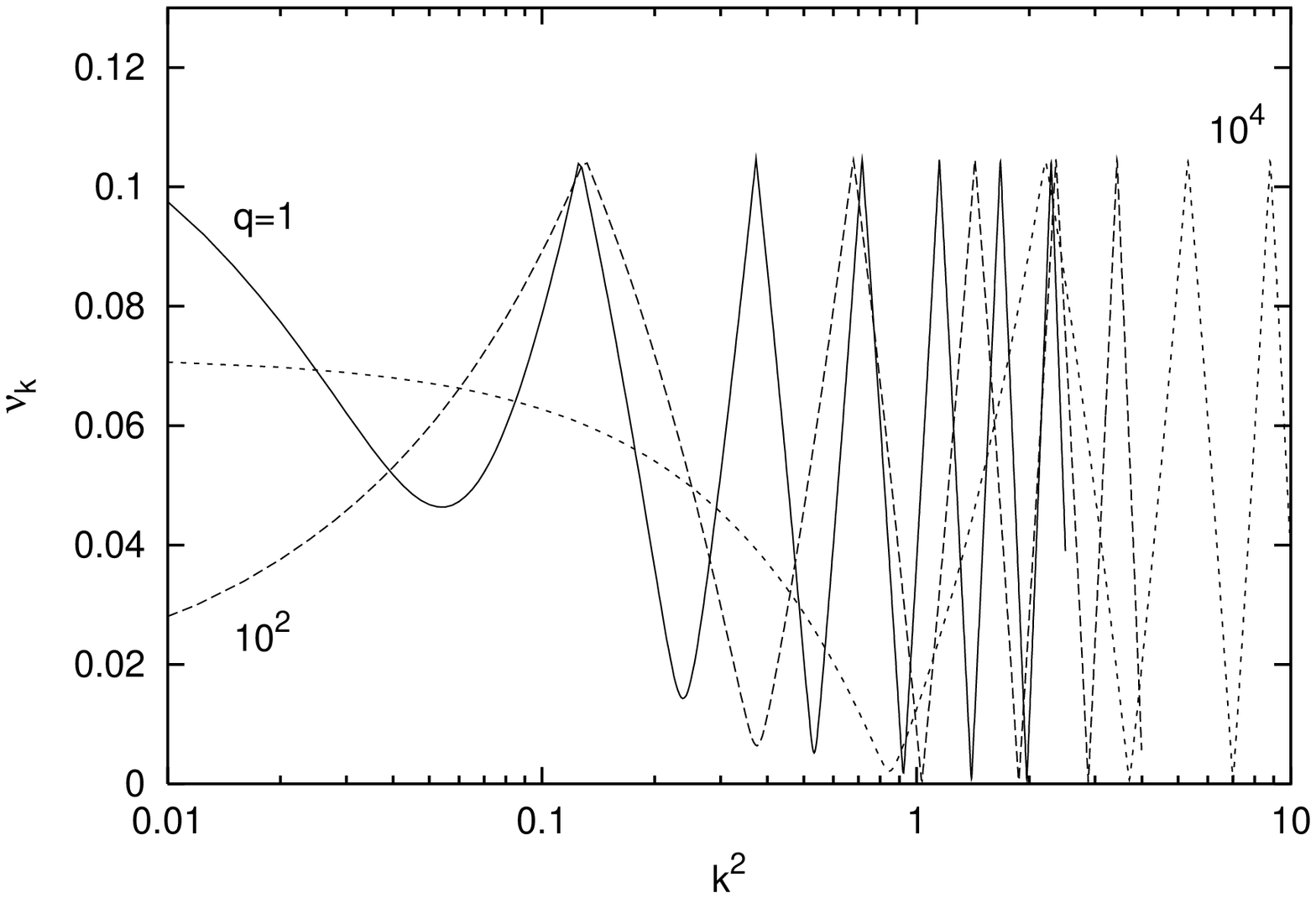,width=8truecm,angle=-90}
\caption{ The fermion (envelope) spectrum $F_k$ (upper panel)
and $\nu_k$ (lower panel) as a function of 
$k^2$, for fermions coupled to the Higgs, with 
$q_2=1, 10^2, 10^4$. Note how narrow the peaks become for
increasing values of $q_2$.}
\label{fknukh}
}

The expressions for $F_k$ and $\nu_k$ are given by Eqs.~(\ref{Fk})
and~(\ref{nuk}), respectively. In this case the function with smaller
frequency in the peaks of $F_k$ corresponds to the solution of
Eq.~(\ref{nuk}) with the plus sign choice. We show in Fig.~\ref{insth}
the instability chart for fermions coupled to the Higgs. The resonance
bands are narrower than those corresponding to the coupling to the
inflaton.

The upper panel in Fig.~\ref{fknukh} shows $F_k$ as a function of $k^2$
for several values of $q_2$, between 1 and $10^4$, while the lower panel 
shows the corresponding
$\nu_k$. Note how quickly the bands become very narrow as $k^2$ is
increased. As a consequence, the fermion production through the coupling 
to the Higgs is in general less efficient than through the coupling to the 
inflaton in hybrid inflation models.

The ratio of Higgs-coupled fermion energy density to total energy is 
given by
\begin{equation}\label{ratiorhohiggs}
{\rho_{_F}(\tau)\over \rho_{\rm total}}= {g^2h_2^2\over\lambda}\,
{2N\over\pi^2 q_2}\int dk\,k^2\,\Omega_k(\tau) \bar n_k(\tau)\,.
\end{equation}
and it is shown in Fig.~\ref{rhohvst}. 
If we take as a crude fit of the numerical results
that $(\lambda /g^2 h_2^2) {\rho_{_F}}/{\rho_{\rm total}} \sim
10^{-2}/q_2$, we see that $\rho_{_F}/\rho_{\rm total} \sim 10^{-2}
g^4/\lambda$, and since we are working in the regime $\lambda \gg g^2$,
we see that the fraction of energy transferred to fermions here remains
small (but not necessarily negligible).

\FIGURE{\epsfig{file=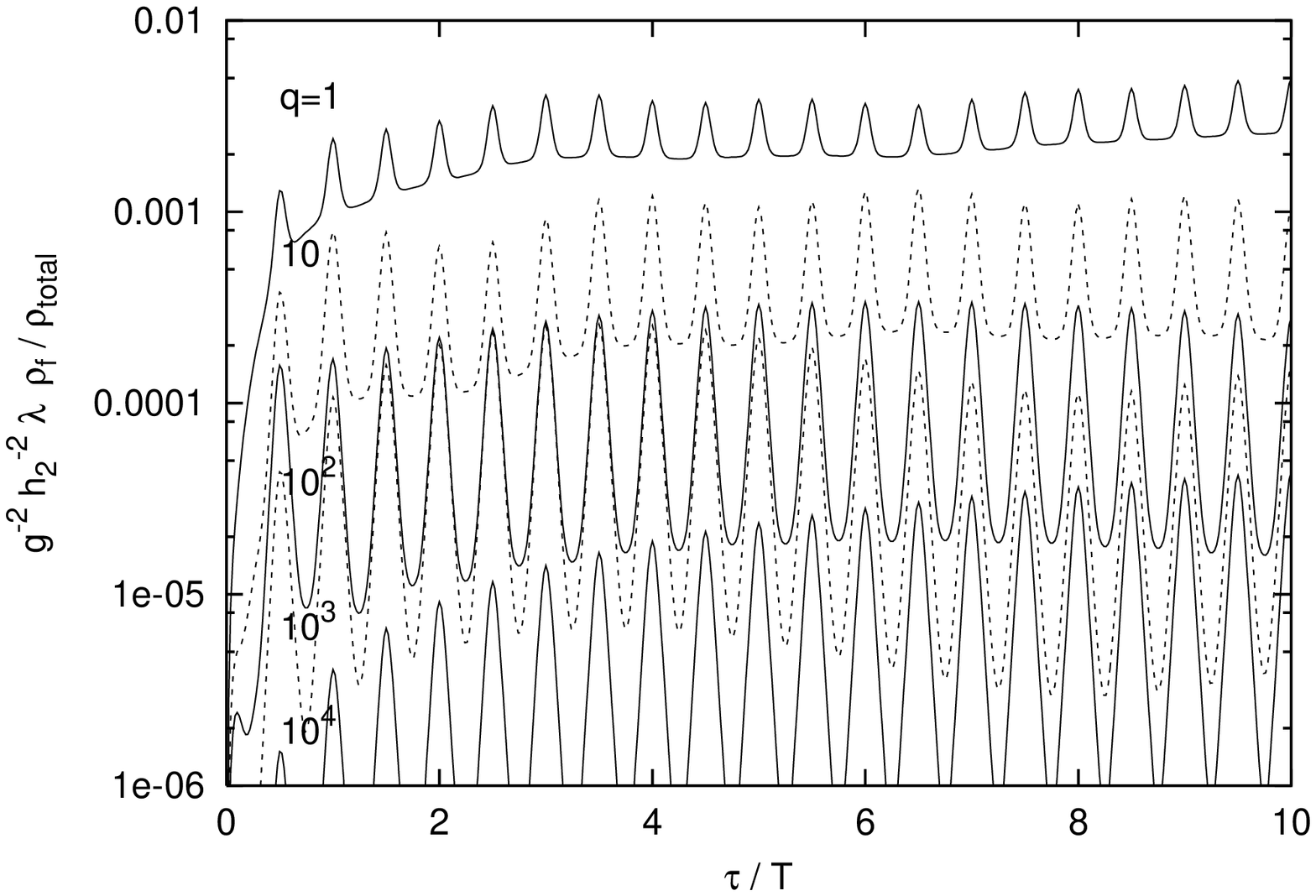,width=8truecm,angle=-90}
\caption{ The fraction of the total energy transferred to fermions
coupled to the Higgs, as a function of time, for $q_2=1, 10, 10^2, 
10^3, 10^4$.}
\label{rhohvst}
}

Since the fermionic production
takes place when $x\simeq 0$, it is necessary to take the initial
condition $y_0$ sufficiently close to one so that small Higgs values
occur. Taking, as we did, $y_0=0.9999$, ensures that at least $x\sim
10^{-2}$ is reached, and hence the parametric resonance is not inhibited
by a mass gap for the values of $q_2$ considered. 

Notice that when the symmetry is finally broken, the fermions get a mass 
$m_\psi=h_2 M/\sqrt{\lambda}$, which also in this case can be quite large.

\section{Boson production}

The parametric resonant production of Higgs field particles 
has been shown to be quite inefficient in hybrid inflation models 
\cite{GBL}. However, other scalar particles coupled to the inflaton 
or the Higgs can can be abundantly produced. 
Here we will compute the parametric production of these bosons in order to 
compare it with the fermionic production.

\FIGURE{\epsfig{file=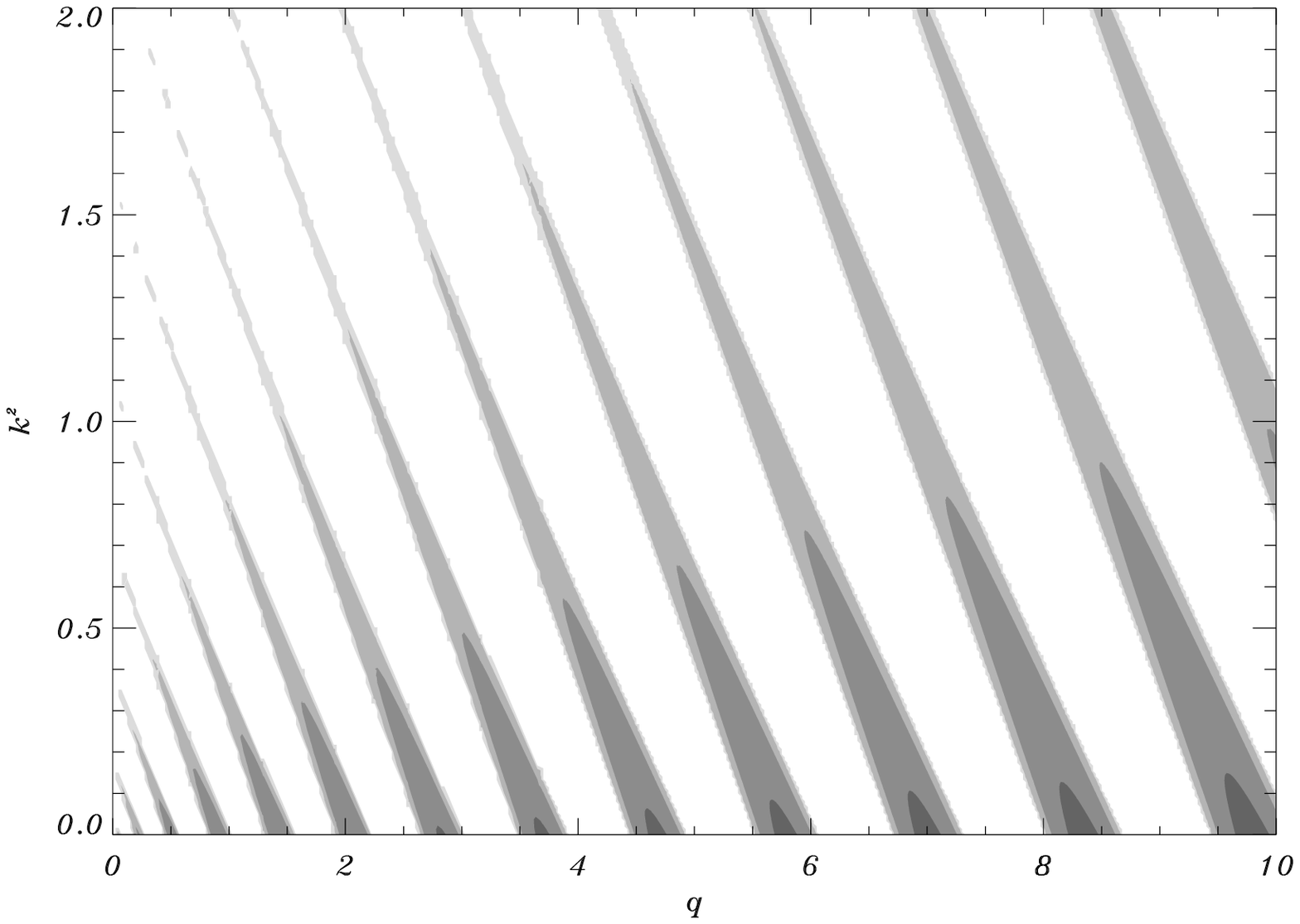,width=12truecm,angle=0}
\epsfig{file=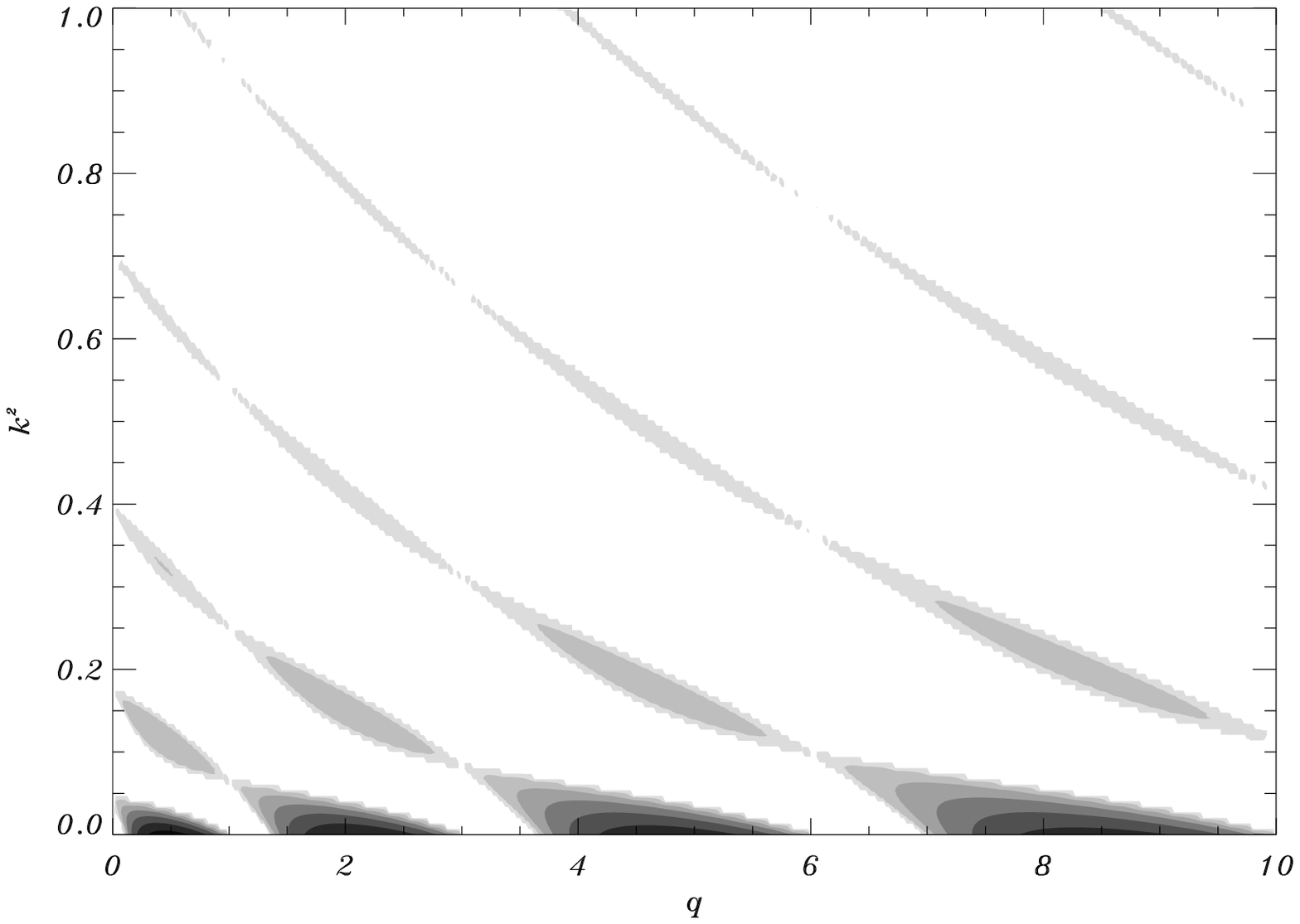,width=12truecm,angle=0}
\caption{ The instability chart for bosons coupled to the
inflaton (upper panel) and to the Higgs (lower panel). 
Contours correspond to equipotential values of 
$\mu_k=10^{-3}$, 0.01, 0.03, 0.05, 0.07 and 0.09  in the plane 
$(q'_{1(2)}, k^2)$.}
\label{instb}
}

We consider a scalar field $\chi$ with mass $m_\chi$ and coupling to 
$\phi$ and $\sigma$ given by
\begin{equation}
V(\chi)= \frac{1}{2}m_\chi^2 \chi^2+\frac{1}{2}g_1^2 \chi^2 \sigma^2
+ \frac{1}{2}g_2^2 \chi^2 \phi^2.
\end{equation}
We can write the equation of motion for the scalar field modes $X_k$ as
\begin{equation}\label{matbos}
X_k''+\Omega_k^2(\tau)X_k(\tau)=0\,,
\end{equation}
where 
\begin{eqnarray}\label{omebos}
\Omega_k^2(\tau)&=&k^2+\bar m_\chi^2+ q_1'\,y^2(\tau) +
q_2'\,(1-y^2(\tau)) \,,\\ 
q_1'&=&{\lambda g_1^2\over g^4}\,,\hspace{1cm}
q_2'={g_2^2\over g^2}\,,\hspace{1cm} 
\bar m_\chi={m_\chi\over\bar M}\,.
\end{eqnarray}

We chose initial conditions of positive frequency plane waves at 
$\tau < 0$
\begin{eqnarray}\label{ic1bos}
X_k(0)&=&[2\Omega_k]^{-1/2}\,,\\
X_k'(0)&=&-i\Omega_kX_k(0)\,.\label{ic2bos}
\end{eqnarray}
The boson occupation number can then be calculated as
\begin{equation}\label{ocnobos}
n_k(\tau)={1\over2\Omega_k}|X'_k|^2 + {\Omega_k\over2}|X_k|^2 -
{1\over2}\,.
\end{equation}
Note that $n_k(0)=0$, thanks to Eqs.~(\ref{ic1bos}) and  (\ref{ic2bos}).

\FIGURE{\epsfig{file=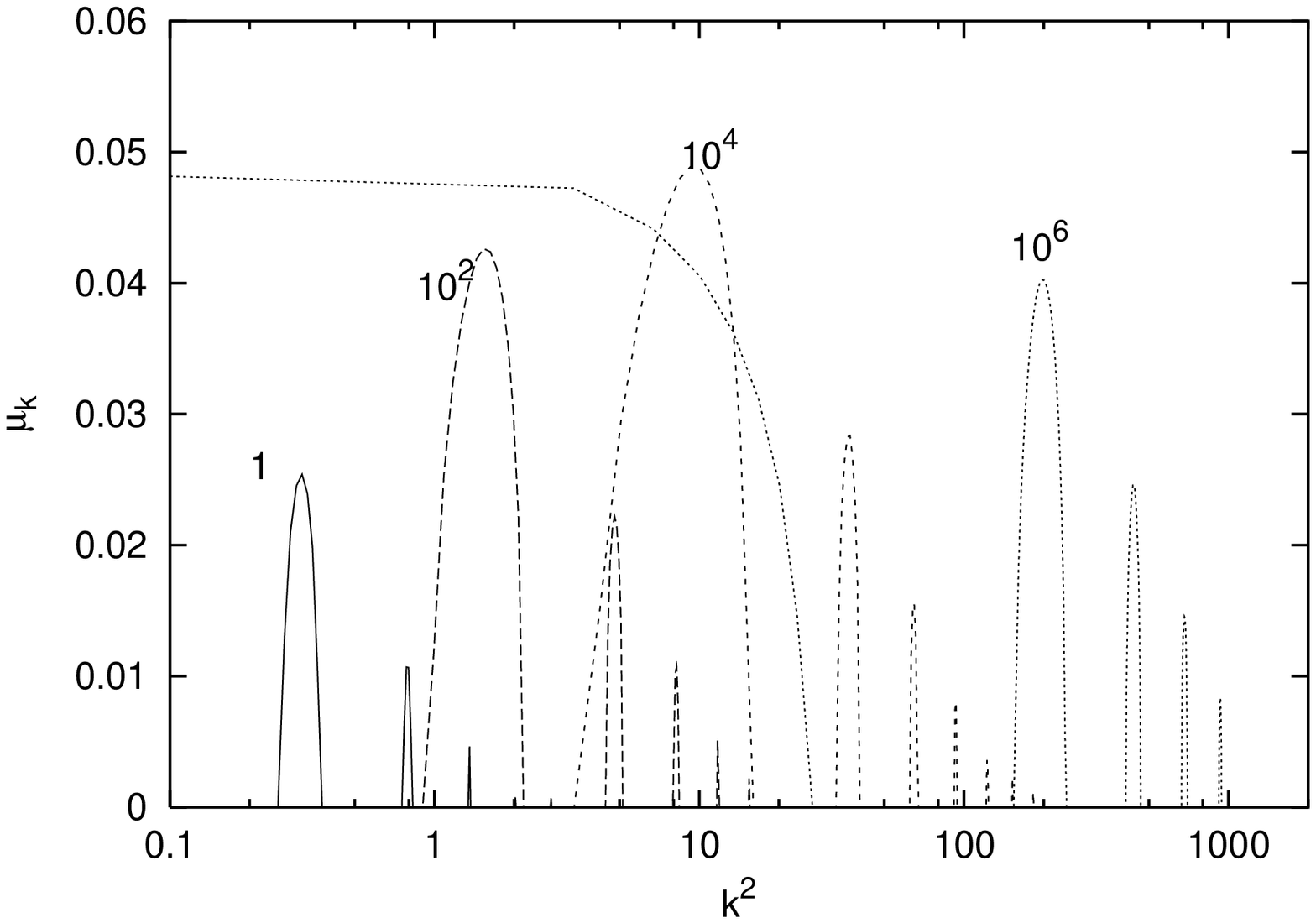,width=8truecm,angle=-90}
\caption{The boson growth index $\mu_k$ as a function of mode number
$k^2$, for bosons coupled to the inflaton. The curves are labelled by
the value of its $q'_{1}$ parameter. Note that the bands become wider
with larger $q'_1$.}
\label{muki}
}

As in the case of the fermion production, we will use the method of 
Ref.~ \cite{GMM} to compute an approximate expression for the boson 
occupation number
\begin{equation}\label{barnkbos}
\bar n_k(\tau)=2\sinh^2(\mu_k\tau)\,,
\end{equation}
where the Floquet index or growth factor $\mu_k$ is determined by:
\begin{equation}\label{muk}
\cosh(\mu_k T)={\rm Re}\,X_k^{(1)}(T)\,,
\end{equation}
with $X_k^{(1)}$ satisfying the same equation (\ref{matbos}) with the
initial condition $X_k^{(1)}(0)=1$, ${X_k^{(1)}}'(0)=0$.
The energy density of bosons can be obtained using this approximate 
solution as
\begin{equation}\label{energybos}
\rho_{_B}(\tau) = 
{1\over2\pi^2}\int dk\,k^2\,\Omega_k(\tau) \bar n_k(\tau)\,.
\end{equation}
and therefore, the ratio of boson to total energies is
\begin{equation}\label{ratiobos}
{\rho_{_B}(\tau)\over \rho_{\rm total}}= {4 g_{1(2)}^2\over \pi^2
q_{1(2)}}\int dk\,k^2\,\Omega_k(\tau) \sinh^2(\mu_k\tau)\,.
\end{equation}

We show in Fig. \ref{instb} the instability chart for bosons coupled to
the inflaton (upper panel) and to the Higgs (lower panel). Shaded areas
correspond to the instability bands for boson production. The darker
areas correspond to larger Floquet index $\mu_k$, and thus to a more
efficient particle production. In the unshaded areas there is no
exponential boson production. The instability bands are thinner in the
case of the coupling to the Higgs compared to those corresponding to the
coupling to the inflaton, and they shrink to zero for specific values
$q'_2= 1, 3, 6, 10,$ etc. This reflects the fact that modes with
$q'_2=n(n+1)/2$ have no instabilities~\cite{JGB}.

\FIGURE{\epsfig{file=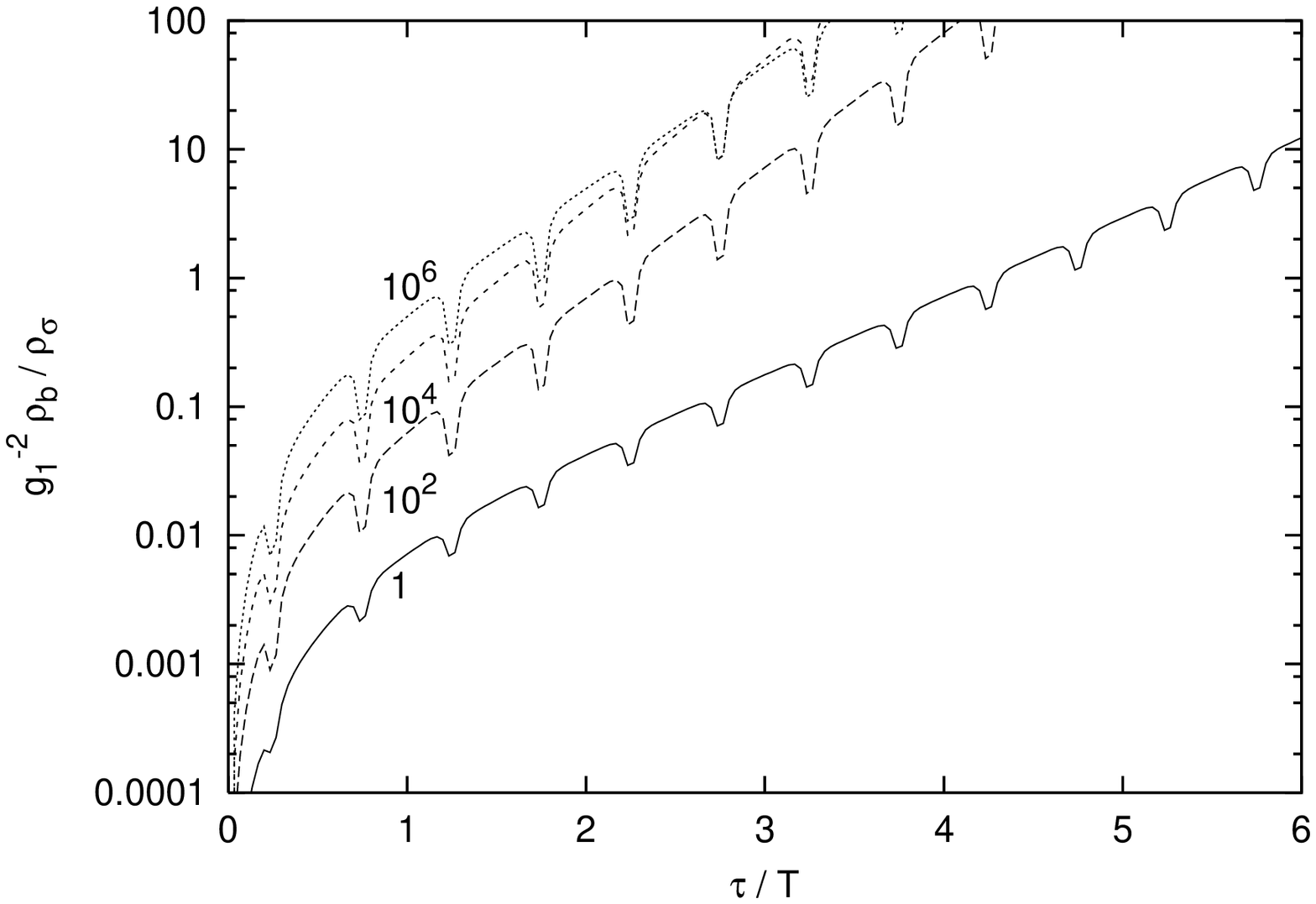,width=8truecm,angle=-90}
\caption{ The fraction of the total energy transferred to bosons
coupled to the inflaton, as a function of time, for different values of
the $q'_1 = 1, 10^2, 10^4, 10^6$ (labelling the curves).}
\label{figrhob}
}

In Fig.~\ref{muki} we show the Floquet index $\mu_k$ for bosons coupled
to the inflaton and different values of the parameters $q'_{1}$ up to
$10^6$, corresponding to vertical slices in the instability charts of
Fig.~\ref{instb} at those $q'_1$ values. The instability bands become
wider and extend up to larger $k^2$ with increasing $q'_1$ values. 
Fig.~\ref{figrhob} shows the fraction of the total energy
transferred to bosons coupled to the inflaton. Its mean value grows
exponentially with time. The exponent is larger for larger $q'_1$
values and the boson production is more efficient.  
When this fraction approaches unity, the backreaction effect
becomes important and the results will be modified~\cite{BHP}. These plots
have to be compared with those in Figs.~\ref{rhoivst} and \ref{rhohvst}
for fermions coupled to the inflaton and Higgs respectively. It is clear
that during the first few oscillations of the inflaton, the energy
transfer to fermions can be more important than that to bosons.

\FIGURE{\epsfig{file=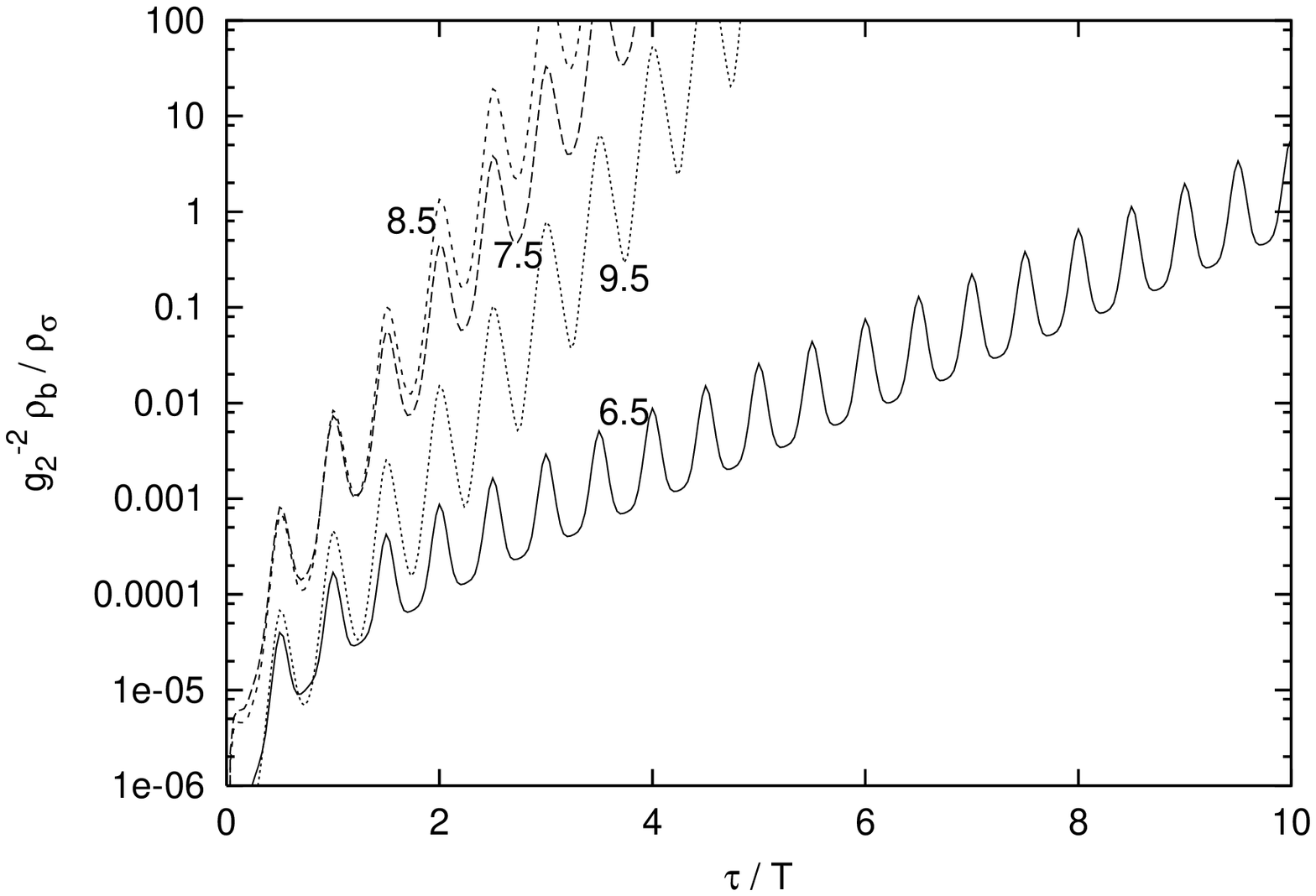,width=8truecm,angle=-90}
\caption{ The fraction of the total energy transferred to bosons
coupled to the Higgs as a function of time for different values of
the $q'_2=6.5, 7.5, 8.5, 9.5$ parameter (labelling the curves).}
\label{rhobh}
}

In the case that the bosons are coupled to the Higgs, as we noticed
above, the particle production vanishes for $q'_2=n(n+1)/2$, and thus no
energy is transferred to fermions for those $q'_2$ values. We show in
Fig.~\ref{rhobh} the fraction of the total energy transferred to
bosons as a function of time for a set of $q'_2$ values spanning the
range between the two zeroes, corresponding to $n=3$ and 4. 
The efficiency of
bosonic production has an oscillatory behaviour with increasing $q'_2$:
it is maximal for $q_2' \sim n^2/2$ and minimal (zero) for $q'_2=n(n+1)/2$.

\section{Conclusions}

In this paper we have studied the parametric resonant production of
fermions in hybrid inflation, with both fields, inflaton and Higgs
coherently oscillating after inflation. We have assumed that fermions
may couple to either the inflaton or the Higgs (or both). The behaviour
in the two cases is very different. While fermion production is very
important in the case of a coupling to the inflaton, even in the
presence of a bare mass, the production of fermions coupled only to the
Higgs is generically weak. This is related to the fact that the
non-adiabaticity condition, $d\Omega_k/d\tau >\Omega_k^2$, is harder to
achieve for the Higgs since, when $\Omega_k$ is at a minimum,
$d\Omega_k/d\tau$ is also at a minimum (contrary to the inflaton
case). When the bare fermion mass exceeds the value $\bar m_\psi
>\sqrt{q_1}$ the fermion production by the coupling to the inflaton
field is also suppressed.

We have studied the growth of the fermion energy density and seen that
it very quickly saturates to an approximately constant value. For
fermions coupled to the inflaton, the asymptotic value grows with the
resonance parameter like $h_1^2 q_1^{1/4}$. For natural values of the
couplings, a significant fraction of the inflaton energy can be
transferred to fermions.  On the other hand, for fermions coupled to the
Higgs, the fermion energy density is of the order $\rho_{_F}/\rho_{\rm
total} \sim 10^{-2}g^4/\lambda$, which under our working assumptions
($g^2\ll\lambda$) is quite small.

We have also studied the boson production for both couplings to inflaton
and Higgs, and compared with the fermion production. While the energy 
density transferred to the parametrically produced fermions saturates
after a few oscillations, the one in bosons grows exponentially with time.
Hence, if both fermions
and bosons have similar couplings, most of the particle production goes
initially into fermions, while at late times the boson production is
exponentially dominant. 
There are however some values of boson couplings to the Higgs,
corresponding to $q'_2=n(n+1)/2$,  which completely
inhibit the parametric resonance of bosons. 

When the energy density of the bosonic or fermionic particles produced 
becomes sizeable, they are expected to backreact on the inflaton, 
affecting its evolution and eventually suppressing the parametric 
production of particles. We have not considered this process in 
detail in this paper since the approach followed is not the most 
suited one for this purpose. For a proper discussion of this issue
in the context of chaotic inflation, see Refs.~\cite{GRTP,BHP}.

As a summary, we have shown that the production of fermions in the 
preheating stage of hybrid inflation can be very important. 
For the range of model parameters assumed, hybrid inflation models
lead to a more efficient fermion production than chaotic inflation models
\cite{GRTP}, without the need to go to extremely large values of the 
resonance parameter $q$.  Depending 
on the relative size of the couplings, and on the backreaction process, 
the inflaton energy transferred to fermions may even be larger than that 
transferred to bosons. At any rate, the parametric production of out 
of equilibrium fermions could have interesting consequences for 
cosmological issues such as the generation of the baryon asymmetry
through e.g. the leptogenesis mechanism.

Note added: while completing the writing of this work a related paper
\cite{BGM} appeared where the gravitino production during preheating is 
computed in a supersymmetric model of hybrid inflation.

\section*{Acknowledgements}

The work of JGB has been supported in part by a CICYT project
AEN/97/1678. SM and ER acknowledge financial support from CONICET,
Fundaci\'on Antorchas and Agencia Nacional de Promoci\'on Cient\'\i fica
y Tecnol\'ogica. SM and ER thank Centro At\'omico Bariloche for
hospitality during the completion of this work.


\begin{thebibliography}{99}

\bibitem{KLS} L. Kofman, A.D. Linde and A.A. Starobinsky, Phys. Rev.
Lett. {\bf 73} (1994) 3195 [hep-th/9405187]; Phys. Rev. D {\bf 56}
(1997) 3258 [hep-ph/9704452].

\bibitem{GKLS} P.B. Greene, L. Kofman, A.D. Linde and A.A. Starobinsky,
Phys. Rev. D {\bf 56} (1997) 6175 [hep-ph/9705347].

\bibitem{GBL} J. Garc\'{\i}a-Bellido and A.D. Linde, Phys. Rev. D
        {\bf 57} (1998) 6075 [hep-ph/9711360].

\bibitem{GKS} P.B. Greene, L. Kofman and A.A. Starobinsky,
Nucl. Phys. {\bf B 543} (1999) 423 [hep-ph/9808477].

\bibitem{GK} P.B. Greene and L. Kofman, Phys. Lett. {\bf B 448}
(1999) 6 [hep-ph/9807339].

\bibitem{BHP} J. Baacke, K. Heitmann and C. Patzold,
Phys. Rev. D {\bf 58} (1998) 125013 [hep-ph/9806205].

\bibitem{GRTP} G. F. Giudice, A. Riotto, I. Tkachev and M. Peloso,
JHEP {\bf 08} (1999) 014 [hep-ph/9905242].

\bibitem{GUTB} E.W. Kolb, A.D. Linde and A. Riotto, Phys. Rev. Lett.
{\bf 77} (1996) 4290 [hep-ph/9606260]; 
G.W. Anderson, A.D. Linde and A. Riotto, Phys. Rev. Lett.
{\bf 77} (1996) 3716 [hep-ph/9606416].

\bibitem{EWB} J. Garc\'{\i}a-Bellido, D. Grigoriev, A. Kusenko and 
M. Shaposhnikov, Phys. Rev. D {\bf 60} (1999) 123504 [hep-ph/9902449].

\bibitem{AS} G. Aarts and J. Smit, Nucl. Phys. {\bf B 555} (1999) 355
[hep-ph/9812413]; Phys. Rev. D {\bf 61} (2000) 025002 [hep-ph/9906538].

\bibitem{hybrid} A.D. Linde, Phys. Lett. {\bf B259} (1991) 38; 
Phys. Rev. D {\bf 49} (1994) 748 [astro-ph/9307002].

\bibitem{GBW} J. Garc\'{\i}a-Bellido and D. Wands, Phys. Rev. D
        {\bf 54} (1996) 7181 [astro-ph/9606047].

\bibitem{CLLSW} E.J. Copeland, A.R. Liddle, D.H. Lyth, E.D. Stewart 
and D. Wands, Phys. Rev. D {\bf 49} (1994) 6410 [astro-ph/9401011].

\bibitem{GBLW} J. Garc\'{\i}a-Bellido, A.D. Linde and D. Wands, 
Phys. Rev. D {\bf 54} (1996) 6040 [hep-ph/9605094].

\bibitem{LR} For a recent review, see
D.H. Lyth and A. Riotto, Phys. Rept. {\bf 314} (1999) 1
[hep-ph/9807278].

\bibitem{Lyth} D.H. Lyth, hep-ph/9904371.

\bibitem{BKS} M. Bastero-Gil, S.F. King, J. Sanderson,
Phys. Rev. D {\bf 60} (1999) 103517 [hep-ph/9904315].

\bibitem{GMM} A.A. Grib, S.G. Mamaev and V.M. Mostepanenko,
{\em Quantum effects in strong external fields}, Friedmann 
Laboratories Pub. Co., San Petersburg (1994).

\bibitem{JGB} J. Garc\'{\i}a-Bellido, in preparation.

\bibitem{BGM} M. Bastero-Gil and A. Mazumdar, hep-ph/0002004.

\end{thebibliography}
\end{document}